\documentclass[sigconf, nonacm]{acmart}
\usepackage{placeins}
\AtBeginDocument{%
  }
\usepackage{stfloats}
\begin{document}
\title{Recommendation Feedback Loops Shape Urban Income Segregation and Venue Visitation}
\author{Guangyuan Weng}
\authornote{Both authors contributed equally to this research.}
\affiliation{%
  \department{Network Science Institute}
  \department{Khoury College of Computer Sciences}
  \institution{Northeastern University}
  \city{Boston}
  \state{MA}
  \country{USA}}
\email{weng.g@northeastern.edu}

\author{Lin Chen}
\authornotemark[1]
\affiliation{%
  \department{Network Science Institute}
  \department{Department of Physics}
  \institution{Northeastern University}
  \city{Boston}
  \state{MA}
  \country{USA}}
\email{l.chen2@northeastern.edu}

\author{Ciro Beneduce}
\affiliation{%
  \institution{Fondazione Bruno Kessler}
  \department{Department of Information Engineering and Computer Science}
  \institution{University of Trento}
  \city{Trento}
  \country{Italy}}
\email{cbeneduce@fbk.eu}

\author{Giovanni Mauro}
\affiliation{%
  \institution{ISTI-CNR}
  \city{Pisa}
  \country{Italy}}
\email{giovanni.mauro@isti.cnr.it}

\author{Massimiliano Luca}
\affiliation{%
  \institution{Fondazione Bruno Kessler}
  \city{Trento}
  \country{Italy}}
\email{mluca@fbk.eu}

\author{Takahiro Yabe}
\affiliation{%
  \department{Tandon School of Engineering}
  \institution{New York University}
  \city{New York}
  \state{NY}
  \country{USA}}
\email{takahiroyabe@nyu.edu}

\author{Bruno Lepri}
\affiliation{%
  \institution{Fondazione Bruno Kessler}
  \city{Trento}
  \country{Italy}}
\email{lepri@fbk.eu}

\author{Luca Pappalardo}
\affiliation{%
  \institution{ISTI-CNR}
  \institution{Scuola Normale Superiore}
  \city{Pisa}
  \country{Italy}}
\email{luca.pappalardo@isti.cnr.it}

\author{Esteban Moro}
\affiliation{%
  \department{Network Science Institute}
  \department{Department of Physics}
  \institution{Northeastern University}
  \city{Boston}
  \state{MA}
  \country{USA}}
\email{e.moroegido@northeastern.edu}
\renewcommand{\shortauthors}{Weng et al.}
\begin{abstract}%
Location-based recommender systems shape where people go in cities and learn from the visits they steer, forming a human--AI feedback loop. 
This feedback loop shapes whom people share places with across income groups and which venues keep attracting customers. 
Yet platforms judge these systems by their offline accuracy on past visits, a test in which the suggestions never change where people go and which therefore cannot capture long-term effects.
Studying these effects in cities is challenging because, unlike online resources, travel constraints and limited venue capacity shape where people can go, while each recommended visit also changes whom other visitors encounter.
To capture such features, we build a long-term simulation framework based on fine-grained human mobility data from four U.S. metropolitan areas, with distance and capacity constraints. 
Focusing on food venues, we study two urban consequences: how unevenly people’s exposure is distributed across income groups (experienced income segregation) and the share of venues receiving no visits from simulated residents (venue disappearance).
We find that, under every algorithm and in every metropolitan area, at least a fifth of venues stop receiving visits, and that more than half of this loss arises from retraining.
More importantly, more accurate algorithms leave people more segregated, and the most accurate algorithm leaves people 15--21\% more segregated than they actually are.
We then test a re-ranking intervention that favours less popular venues and runs on top of a platform's existing algorithm.
It lowers both venue disappearance and segregation across algorithms, largely without directing people to venues where they may feel less comfortable.
Our results suggest that impact assessments of recommender systems should account for the feedback loop and report segregation and venue disappearance beyond accuracy.

\end{abstract}
\maketitle

\section{Introduction}\label{sec:intro}
Online platforms shape which places people discover and visit in urban environments~\cite{anderson2012learning}: maps and review platforms rank the restaurants nearby, food apps suggest where to eat next, and large language models (LLMs) answer open-ended questions about where to go~\cite{banerjee2020analyzing,pappalardo2024survey,chen2026large, cornacchia2026traffic}.
Unlike a printed guide, the location-based recommender systems behind these online platforms learn from the visits they steer, forming a human--AI feedback loop that affects both people and places~\cite{pedreschi2025human}.
For people, a crucial outcome at stake is \emph{experienced income segregation}: how unevenly an individual's time at venues spreads across the income groups~\cite{moro2021mobility,athey2021estimating,nilforoshan2023human,xu2025using}.
Low experienced income segregation means a person's daily places mix incomes, creating chances for cross-income ties, which predict upward income mobility~\cite{chetty2022social}. 
For places, a key outcome at stake is \emph{venue disappearance}: the share of venues that receive no visits from simulated residents. 
The feedback loop can concentrate visits on already popular venues~\cite{mauro2026urban}, and a venue without visits is then no longer recommended. 
Venue disappearance matters for small businesses that depend on online visibility to attract customers~\cite{banerjee2020analyzing}, as well as for the neighbourhoods they serve.
As the feedback loop unfolds and the recommender system keeps retraining, both outcomes can build up over time, and their burden can fall unevenly across income groups and neighbourhoods.
These effects of the feedback loop are distributional, long-term consequences of algorithmic deployment on the Web. 
They extend beyond the screen into residents' lived experiences of the city and bear on the fairness of the urban environments that online platforms increasingly help to shape. 
Understanding and mitigating them is therefore integral to building a responsible Web~\cite{milano2020recommender}.

Yet these algorithms are rarely evaluated for their long-term effects on communities, including how they shape segregation across income groups and the distribution of visits across local businesses. Standard offline accuracy measures assess how well an algorithm predicts past behaviour, leaving these collective outcomes largely unexamined.
This \emph{out-of-loop} test evaluates the recommender system once, using data describing visits that were not shaped by its recommendations. 
Therefore, it cannot capture experienced income segregation or venue disappearance that may emerge as the feedback loop unfolds~\cite{damour2020fairness,mcnee2006being}. 
Although regulation has begun to require platforms to assess the systemic risks of their recommender systems~\cite{eu2022dsa}, methods for assessing such long-term effects are still lacking.

Prior work has mainly studied human--AI feedback loops for online items such as songs, movies and products~\cite{jiang2019degenerate,chaney2018algorithmic}, which any number of users can consume without affecting one another.
Urban venues, by contrast, differ from online items in at least three respects:
\textbf{(1) Spatio-temporal constraints.} People choose venues near their current location and at times that fit their daily routines; recommendations therefore shape choices within a small set of locally accessible options.
\textbf{(2) Finite capacity.} A restaurant can serve only a limited number of customers at a time; when recommendations send many people to the same venue, some must wait or go elsewhere.
\textbf{(3) Social coupling.} Visitors to a venue share the same physical space; as a result, a recommendation to an individual also changes whom the other visitors meet.

To the best of our knowledge, no existing study can capture these three factors together. Audits find that review sites and LLM assistants expose venues unevenly across neighbourhoods~\cite{singhal2025auditing,chen2026large}, but they do not measure how exposure changes as recommendations and user behaviour feed back on one another. Simulations can reproduce the feedback loop, but they simplify the setting: some use synthetic users~\cite{chaney2018algorithmic}, others focus on online content or retail goods~\cite{anwar2024filter,barlacchi2026diversity}. 
The closest urban simulation study measures how venue concentration changes throughout the feedback loop but does not account for venue capacity nor measure experienced income segregation~\cite{mauro2026urban}. 
Taken together, these gaps leave three research questions unanswered:
\begin{itemize}
  \item \textbf{RQ1:} How does retraining within the feedback loop affect experienced income segregation and venue disappearance?
  \item \textbf{RQ2:} How do different recommendation algorithms shape experienced income segregation and venue disappearance?
  \item \textbf{RQ3:} Can an algorithmic intervention reduce venue disappearance and experienced income segregation within the feedback loop?
\end{itemize}

\begin{figure}[t]
\centering
\includegraphics[width=\columnwidth]{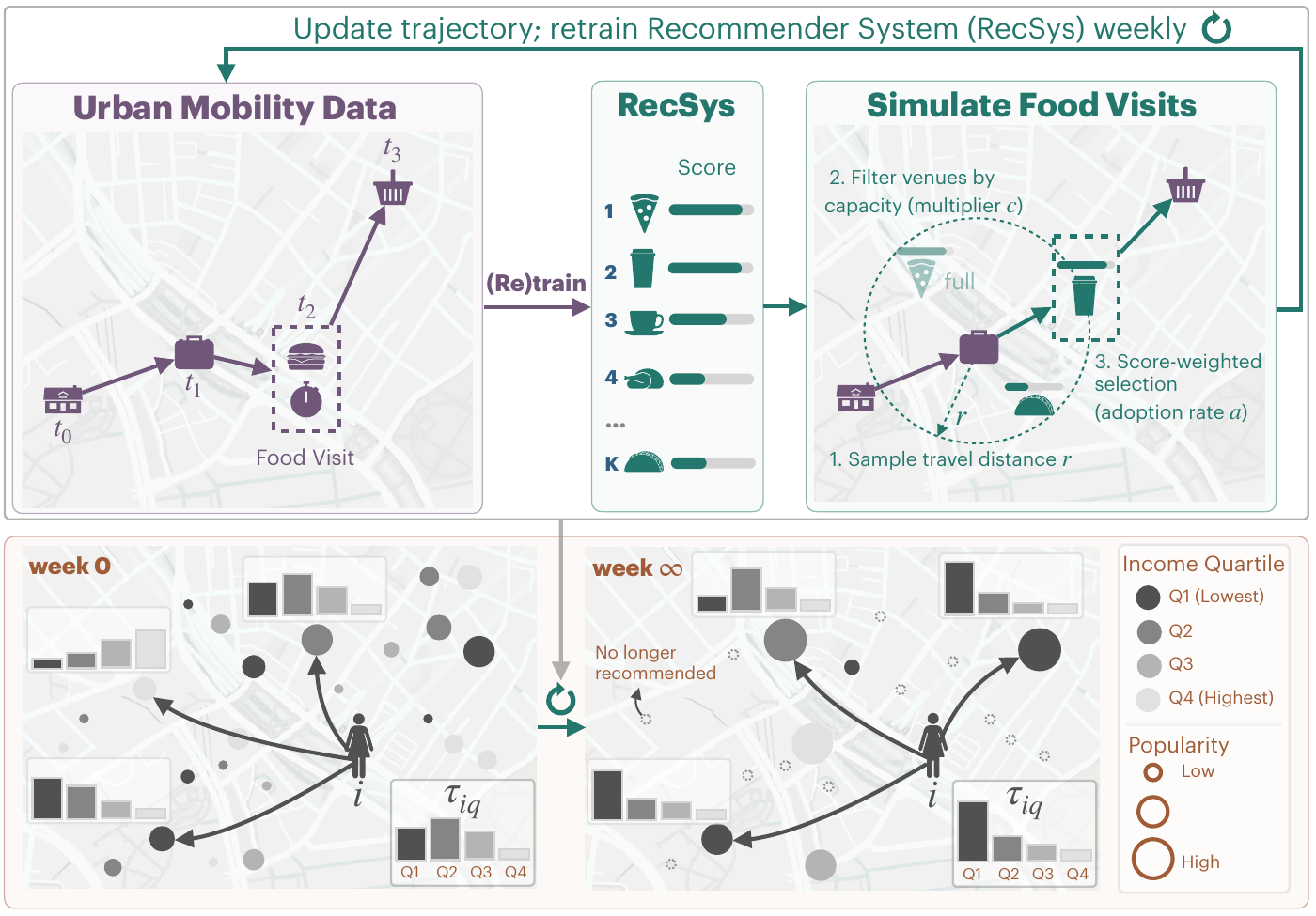}
\Description{Framework diagram in three linked parts: mobility data, simulated food visits with a recommender system, and outcome maps. Top left, Urban Mobility Data: on a street map, a resident travels from home at t0 to work at t1, to a burger venue at t2, and to a shop at t3. A purple dashed box labelled Food Visit encloses the burger and a stopwatch for the visit's duration. An arrow labelled (Re)train leads to RecSys, which ranks food venues from 1 to K with score bars: pizza, takeaway cup, coffee cup, chicken, and a taco at rank K. An arrow leads on to Simulate Food Visits. There, step 1 samples a travel distance r around the resident's last venue, drawn as a dashed circle. Step 2 filters venues by capacity with multiplier c, so the top-ranked pizza venue, marked full, is skipped. Step 3, score-weighted selection with adoption rate a, picks the second-ranked takeaway cup, highlighted by a teal dashed box, and the resident continues to the shop. An arrow labelled Update trajectory, retrain Recommender System (RecSys) weekly returns to the data. A downward arrow leads from these panels to a circular retraining symbol between two venue maps, week 0 and week infinity, joined by a rightward arrow. Circles are venues, shaded by income quartile from Q1 (lowest, black) to Q4 (highest, light grey) and sized by popularity. In both maps, arrows lead from a resident i to the three venues they visit, and small bar charts give each visited venue's income mix. At week 0, these venues are dominated by different quartiles, and the inset tau i q, the resident's exposure to the four quartiles, is fairly even. At week infinity, two of the three venues are dominated by Q1, and tau i q is dominated by Q1. Many small circles become dashed, labelled no longer recommended, indicating venue disappearance.}
\caption{Our simulation of the human--AI feedback loop, grounded in real human mobility. On each food trip, the recommender system suggests venues under distance and capacity constraints, and it is retrained weekly on the previous four months of residents’ trajectories. Circles mark venues, and dashed circles are venues no longer recommended (venue disappearance). Experienced income segregation measures how unevenly person $i$'s exposure $\tau_{iq}$, the time-weighted average of the income mixes of the venues they visit (bars), spreads across income quartiles.}
\label{fig:c1}
\end{figure}

To address these questions, we develop a realistic simulation of the human--AI feedback loop in four large U.S. metropolitan areas (Fig.~\ref{fig:c1}). 
In each area, we simulate 4,000--8,000 residents on their GPS-recorded daily schedules, drawn in equal numbers from four income quartiles.
We model recommendations only for food venues and exclude destinations for trips home or to work because they are largely determined by daily routines.
We compare the outcomes of four representative recommendation algorithms:
UR-POI (User Returner over Points Of Interest) recommends the venues a person visits most often, ItemKNN and UserKNN learn from co-visitation patterns through similar venues~\cite{sarwar2001item} and similar people~\cite{resnick1994grouplens}, respectively, and Markov uses population-wide transitions from the last venue~\cite{gambs2012next}. 
To isolate the effect of retraining, we compare our results with those from an out-of-loop baseline, where each algorithm is trained just once on the observed visits. 

Our contributions are as follows:
\begin{itemize}
  \item \textbf{A simulation framework for assessing the long-term effects of location-based recommender systems on people and places, grounded in large-scale mobility data and real-world constraints.}  It builds on residents’ GPS-recorded daily schedules, respects travel distances and venue capacities, and measures both experienced income segregation and venue disappearance over three years of weekly retraining.
  \item \textbf{We reveal that the feedback loop causes most of the venue disappearance and pushes the algorithms apart in segregation (RQ1).} Under every algorithm, a fifth or more of the venues lose \emph{all} visits from simulated residents, and retraining causes more than half of this loss in every metropolitan area. The loss concentrates in less active, lower-income neighbourhoods. 
  \item \textbf{We find a tradeoff between offline accuracy and the resulting segregation, \emph{i.e.}, the more accurate the algorithm, the more segregated people become (RQ2).} The most accurate algorithm leaves people 15--21\% more segregated than they actually are, and the lowest income quartile, already the most segregated, becomes more segregated still. Segregation comes from how the algorithms rank venues rather than from which venues lie within reach, and platforms can therefore change how segregated their users become.
  \item \textbf{Platforms can mitigate these harms with a simple re-ranking step on top of their existing algorithm (RQ3).} At a modest cost in offline accuracy, promoting less popular venues more than halves venue disappearance under every algorithm and lowers income segregation under three of the four, while largely preserving residents' comfortableness.
\end{itemize}

Together, these contributions can inform the design of responsible location-based recommender systems that keep cities socially mixed and their local businesses alive.
Our code is available at \url{https://github.com/SUNLab-NetSI/recsys-city-feedback}.

\begin{figure}[t]
\centering
\includegraphics[width=0.85\columnwidth]{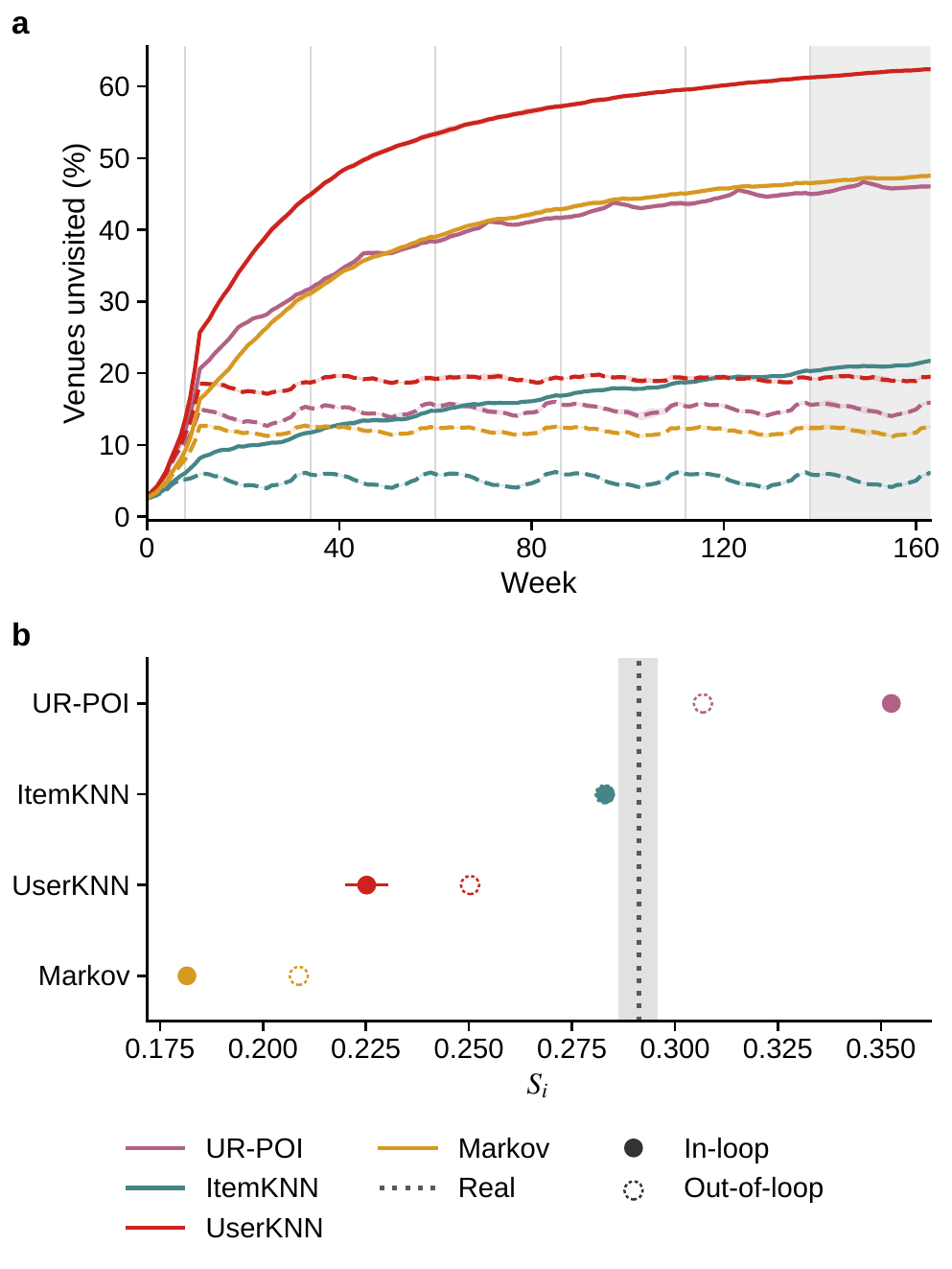}
\Description{Line chart over 164 simulated weeks. Each of four algorithms appears twice in its colour: a solid line for weekly retraining in the loop and a dashed line for training once out of the loop. After the first weeks the dashed lines stay low and nearly flat, while the solid lines keep rising, highest for UserKNN and lowest for ItemKNN. Thin vertical lines mark each repeat of the six-month schedules. A grey column marks the last six months. Below, one row per algorithm: a dashed hollow dot (out of the loop) and a filled dot (in the loop), with no lines between them, with a grey band and dotted line for the observed level.}
\caption{Retraining in the loop drives most of the venue disappearance and pushes the algorithms apart in segregation. Solid lines and filled dots: retrained every week (in the loop); dashed lines and hollow dots: trained once (out of the loop). (a) Share of training venues unvisited in the previous three months; vertical lines: each repeat of the six-month schedules; grey: the last six months. (b) $S_i$ (experienced income segregation) over the last six months; grey band: observed level. Boston, means over five simulations with 95\% intervals.}
\label{fig:c7}
\end{figure}
 
\section{Related Work}\label{sec:related}
\paragraph{Experienced income segregation.}
Segregation is a persistent problem for urban residents, associated with disparities in health~\cite{waitzman1998separate} and with lower upward income mobility~\cite{chetty2014land}.
In recent years, mobility data have enabled the study of experienced segregation, measured in the places people visit every day rather than where they live~\cite{moro2021mobility,weng2025beyond,xu2025using}.
Various studies relate experienced segregation to city size~\cite{nilforoshan2023human}, disaster shocks~\cite{yabe2023behavioral,chen2026urban}, or venue characteristics~\cite{fan2026invisible}, but usually measure it in the movements people actually made.
As experienced income segregation is associated with individuals' mobility patterns~\cite{moro2021mobility}, and location-based recommender systems increasingly steer where people go~\cite{anderson2012learning}, these systems may change whom people meet.
Research on recommender systems has examined how unevenly platforms expose venues across neighbourhoods~\cite{singhal2025auditing} and how recommendations narrow or concentrate what users consume~\cite{anwar2024filter,chaney2018algorithmic,mansoury2020feedback}, but not their effects on experienced income segregation.
It has thus overlooked a critical social consequence that accumulates over time and shapes residents' lived experience of the city.
We address this gap by simulating three years of weekly retraining on real mobility data, and measuring how each recommender system changes whom residents meet across incomes.

\paragraph{Feedback loops in recommender systems.}
A recommender system that retrains on the interactions it has shaped forms a feedback loop.
For each user, such loops narrow the set of items the user is exposed to over time~\cite{jiang2019degenerate, nguyen2014exploring} and amplify the user's preferences toward recommended content~\cite{kalimeris2021preference}.
Across users, they homogenise behaviour without increasing utility~\cite{chaney2018algorithmic}, an effect distinct from the narrowing of each user's consumption~\cite{anwar2024filter}, and they amplify popularity bias through a rich-get-richer effect~\cite{mansoury2020feedback}.
Their net effect on diversity depends on the algorithm and the setting: simulations find that the diversity of each user's consumption can rise while aggregate diversity falls~\cite{fleder2009blockbuster,barlacchi2025simulation,barlacchi2026diversity}, whereas a randomised field experiment on Spotify found the reverse~\cite{holtz2020engagement}.
Nearly all of this evidence concerns online items such as songs, movies and products, which any number of users can consume without affecting one another.
Urban venues differ in three respects: choices are bounded by spatio-temporal constraints~\cite{hagerstrand1970people}, each venue has finite capacity, and, as venues are shared physical spaces, recommendations produce spillover effects on non-recipients~\cite{cornacchia2026traffic} by altering the composition of co-present visitors.
Closest to our work, Mauro et al.~\cite{mauro2026urban} simulate weekly retraining of location-based recommender systems in New York and Tokyo, but the simulation does not model venue capacity, and check-ins, being voluntary, capture only a small share of residents and their visits.
We address these limitations by simulating residents on their GPS-recorded daily schedules in four large U.S. metropolitan areas over three years of weekly retraining under spatio-temporal and capacity constraints, while retaining the observed visits of all other individuals at each venue, so that recommendations to simulated residents also affect co-present non-recipients.

\section{Simulation Framework}\label{sec:framework}
In our simulation, residents follow their observed daily schedules, and a location-based recommender system suggests where they eat on each food trip (Fig.~\ref{fig:c1}).
We restrict our simulation to food trips for two reasons. First, trips that platforms rarely recommend, such as those to home or to work, would mix their effects with those of the recommendations. Second, restaurant visits mix people from different neighbourhoods and are about half as segregated as where people live~\cite{davis2019segregated}, making restaurants an appropriate setting to measure the effect of recommender systems on complex social systems.
The simulation captures the three factors that distinguish urban venues from online items: residents keep their observed schedules of all trips and travel distances, each venue has a finite capacity, and a recommendation to one resident also changes whom the other visitors meet.

\subsection{Data} \label{sec::data}
We use anonymised GPS mobility data provided by Cuebiq, which record people's visits to venues from October 2016 to March 2017.
In each metropolitan area, we simulate residents who made at least 200 recorded visits, visited at least 10 distinct food venues, and visited food venues throughout the six-month observation period. Following the methodology in \cite{moro2021mobility}, we infer each resident's home census block group as the one where they are most often seen between 10 pm and 6 am, rank census block groups within each metropolitan area by their median household income from the American Community Survey five-year estimates~\cite{acs}, and assign each resident the income quartile of their home block group. This provides a neighborhood-level proxy for income rather than a measure of individual income.
We draw equal numbers of residents from the four income quartiles, 4,000 in Boston, San Francisco and Detroit and 8,000 in Chicago.
During the first four months, the training window, these residents visited between 8,946 and 21,852 food venues per metropolitan area. Over the observed six months, their visits cover 95.5--99.4\% of the food venues recorded in each metropolitan area. The last two months are held out for offline evaluation.
Each resident keeps the dates and durations of their observed trips, and we simulate only which venue each food trip goes to. To cover about three years, each resident's schedule repeats every six months.
The other people in the mobility data, between 65,525 and 290,977 per metropolitan area, are not simulated. We keep their real visits unchanged, and they count among each venue's visitors when we measure experienced income segregation (Section~\ref{sec:measures}).

\subsection{Feedback Loop Model} \label{sec:feedback-model}
We develop a simulation of the feedback loop between residents and location-based recommender systems, drawing on previous work~\cite{mauro2026urban} while incorporating large-scale mobility data and realistic constraints on travel behavior and venue capacity. For each food trip, the recommender system suggests venues within reach, and the resident either follows the suggestions or keeps their observed visit.
To keep each trip within a realistic travel distance, we draw a search radius for each trip from the distances that residents travelled to venues they had not visited before during the training window. The recommendations are the $K$ venues that the algorithm ranks highest among those that lie within this radius of the resident's last venue and still have capacity. If no such venue lies within the radius, we draw a new radius.
To account for finite capacity, we give each venue a weekly capacity of $c$ times the residents' observed visits to it in the corresponding week of the observed six months. Once a venue's visits in a simulated week reach this capacity, it is no longer recommended until the next week. Setting $c>1$ allows for busy weeks and for customers willing to wait, while keeping each venue's capacity finite.
We model the probability that a resident follows the recommendations through an adoption rate $a$~\cite{mauro2026urban, cornacchia2026traffic}; otherwise, the resident goes to the venue of their observed visit. A resident who follows the recommendations picks one of the $K$ venues with probability proportional to its min--max scaled score.
At the end of each simulated week, the recommender system is retrained on the residents' trajectories of the previous four months, including the visits that followed its own recommendations, which closes the feedback loop. The simulation runs for about three years.

\subsection{Outcome Measurement} \label{sec:measures}
A location-based recommender system affects both the people who follow its suggestions and the venues they visit. For
people, we measure experienced income segregation, and for venues, venue disappearance. 
Both outcomes are restricted to the training
venues, the food venues that residents visited during the initial four-month training window.
We also report offline accuracy, the measure platforms typically use to evaluate recommender systems. 
Each simulated week, we compute income segregation and venue disappearance using visits from the preceding three months. We report the average of these weekly measures over the final six months of each simulation, when the feedback loop has been running for more than two and a half years.

\subsubsection{Experienced income segregation}
For people, the outcome is whom they share places with across incomes.
People form ties across incomes in the places they share, and such ties are among the strongest predictors of upward
income mobility~\cite{chetty2022social}. 
Mobility data show that whom people meet depends on where they spend their
day as well as on where they live~\cite{moro2021mobility,athey2021estimating,nilforoshan2023human}.
By steering where people eat, a recommender system therefore also shapes whom they can meet across incomes, and with it their chances of moving up the income ladder.
We quantify this using an established measure of experienced income segregation~\cite{moro2021mobility}. 
A person is fully integrated when their
time at venues is split evenly among the four income quartiles of the people around them, and segregated to the extent that their time departs from this even split. 
Let $\tau_{q\alpha}$ be the share of all time spent at venue $\alpha$ by people of income quartile $q$. 
It counts everyone at the venue, including the other people in the mobility data, whose real visits we keep (Section~\ref{sec::data}). 
Let $\tau_{i\alpha}$ be the share of person $i$'s time spent at $\alpha$. 
Person $i$'s exposure to quartile $q$ is $\tau_{iq}=\sum_\alpha \tau_{i\alpha}\tau_{q\alpha}$, and
\begin{equation}
  S_i = \frac{2}{3}\sum_{q=1}^{4}\Big|\tau_{iq}-\frac{1}{4}\Big|,
\end{equation}
which is 0 for an even split and 1 when a person meets only one quartile. Time is weighted by how long people stay. To keep
each estimate reliable, we use only training venues with at least 20 visitors, and we compute a person's
$S_i$ only when they made at least five visits to such venues in the three-month window. 

\subsubsection{Venue disappearance}
For places, the outcome is how many venues lose all their simulated visitors. 
We measure venue disappearance as the share of training venues that no simulated resident visited in the previous three months. 
To capture how unevenly the remaining visits spread, we also report the Gini coefficient of
visits over all training venues.
Venue disappearance matters because many small businesses rely on online platforms to attract customers~\cite{banerjee2020analyzing}.
In the simulation, the recommender system learns only from recent visits and therefore stops recommending venues that receive none.

\subsubsection{Offline accuracy}
Following standard practice, each algorithm is trained once on the training window and then predicts the food visits people made afterwards.
We evaluate these predictions with NDCG@20~\cite{jarvelin2002cumulated}, averaged over each individual's visits and then over people.
Measuring accuracy inside the feedback loop remains an open problem, because the visits to be predicted are themselves shaped by the recommendations, and we therefore report offline accuracy, the measure platforms use to choose an algorithm.
{} \section{Experiment Setup}\label{sec:experiment_setup}
\subsection{Recommendation Algorithms}\label{sec:algorithms}
We compare four recommendation algorithms (Table~\ref{tab:main}), chosen to span the information a recommender system can rely on when ranking venues for a resident, from their own visits (UR-POI), through other people's co-visits (ItemKNN and UserKNN), to the transitions of the whole population (Markov).
\begin{enumerate}
    \item UR-POI ranks the venues a resident has visited by the share of their visits that went to each.
    \item ItemKNN ranks venues by their similarity to the venues the resident has visited, including those venues themselves, where two venues are similar when the same people visit them~\cite{sarwar2001item}.
    \item UserKNN ranks venues by how often they were visited by the $k$ people with the most similar visits, the resident themselves included~\cite{resnick1994grouplens}.
    \item Markov ranks venues by everyone's transitions from the census tract and food type of the resident's last venue~\cite{gambs2012next}.
\end{enumerate}
In Boston, we simulate two additional algorithms for comparison (Appendix~\ref{app:all}): LightGCN~\cite{he2020lightgcn}, a neural model trained on the graph of people and venues, and random choice, which picks uniformly among the reachable venues.

\subsection{Experimental Conditions}\label{sec:conditions}
To isolate the effect of retraining, we add an out-of-loop control as a baseline, in which each algorithm is trained once at the start and never retrained, with everything else unchanged.
As a reference without any recommender system, we also measure each outcome on the residents' observed visits, unchanged, and call these values the observed level.
Offline evaluation (Section~\ref{sec:measures}) limits each prediction to nearby venues, as the simulation does. For each held-out food visit, an algorithm ranks the food venues of the training window that lie within travel distance of the resident's last venue.

To quantify how much of each outcome is due to retraining, we compare each algorithm with its out-of-loop control. 
For venue disappearance, let $U_{\text{in}}$ be the share of training venues left unvisited over the last six months with retraining, and $U_{\text{out}}$ the same share in the control. 
The feedback loop's share of venue disappearance is $(U_{\text{in}}-U_{\text{out}})/U_{\text{in}}$, the fraction of the unvisited share that retraining adds. 
For experienced income segregation, which has an observed level, let $\bar{S}_{\text{in}}$ and $\bar{S}_{\text{out}}$ be the mean $S_i$ over the same period with retraining and in the control, and $\bar{S}_{\text{obs}}$ the mean $S_i$ in the observed visits. 
The feedback loop's share of segregation is
$(\bar{S}_{\text{in}}-\bar{S}_{\text{out}})/(\bar{S}_{\text{in}}-\bar{S}_{\text{obs}})$, the fraction of
the gap from the observed level that retraining creates.

\subsection{Parameters and Replication}
Our main setting uses capacity $c = 1.5$ and $K = 20$ recommendations per trip, with adoption $a = 1$, at which residents always follow the recommendations. ItemKNN and UserKNN use $k = 10$ neighbours~\cite{mauro2026urban}.
Unless noted otherwise, we simulate each setting five times with different random seeds, which change the random draws of adoption, search radius and venue choice and the order in which venues fill up each week. We report means over the five simulations with 95\% confidence intervals from Student's $t$ distribution. For the observed visits, the intervals come from a bootstrap over people.
For robustness, we also vary the capacity $c$ from 1.0 to uncapped and adoption $a$ from .25 to .75 in Boston (Fig.~\ref{fig:wA6}). 
\section{Results}\label{sec:results}
\subsection{Contribution of Retraining to Venue Disappearance and Income Segregation} \label{sec:loop}

To quantify how retraining changes venue disappearance and experienced income segregation, we compare each algorithm in the feedback loop with its out-of-loop control (Section~\ref{sec:conditions}).
In Boston, the share of unvisited venues stabilises within the first weeks without retraining, but continues to grow over the three-year simulation with weekly retraining (Fig.~\ref{fig:c7}a).
Across the four metropolitan areas, over the last six months, 21--71\% of the training venues are unvisited in the loop, against 5--30\% out of the loop.
\emph{In every metropolitan area, retraining accounts for 55--75\% of the venue disappearance under every algorithm.}
It also raises the Gini of visits by 6--15\% relative to the out-of-loop control, least under ItemKNN and most under Markov (Table~\ref{tab:loop}).
This suggests that most of the venue disappearance arises through retraining, rather than from how an algorithm ranks venues when it is first trained.

The loss accumulates as the recommender system is retrained every week on the visits of the previous four months, including those that followed its recommendations.
Venues without visits in this period are no longer recommended.
Offline evaluation tests predictions against held-out observed visits and therefore cannot capture the venue disappearance produced by this feedback loop.

\emph{Retraining also changes whom people share venues with} (Fig.~\ref{fig:c7}b).
Out of the loop, experienced income segregation is already higher than in the observed visits under UR-POI and lower under UserKNN and Markov, and retraining widens both gaps.
In every metropolitan area, it raises $S_i$ under UR-POI by .035 to .046 and lowers it
under UserKNN and Markov, while ItemKNN barely moves (Table~\ref{tab:loop}).
For UR-POI, retraining accounts for 74--79\% of its excess over the observed level.
Retraining therefore pushes the algorithms further apart in segregation, making algorithm choice more consequential than the out-of-loop comparison suggests.

\subsection{Relationship between Offline Accuracy and Income Segregation across Algorithms} \label{sec:people}
To assess the social consequences of selecting algorithms by offline accuracy, we compare their accuracy with the experienced income segregation they produce inside the feedback loop.
\emph{Choosing among the four algorithms by offline accuracy alone selects the most segregating one in every metropolitan area} (Table~\ref{tab:rq1}).
Offline NDCG@20 and experienced income segregation rank the four algorithms in the same order, from UR-POI through ItemKNN and UserKNN to Markov, which is also the order in which they rely more and more on other people's visits.
Compared with residents' observed visits, experienced income segregation is 15--21\% higher under UR-POI, similar under ItemKNN, and lower under UserKNN and Markov.
UR-POI achieves the highest offline accuracy because it exploits the tendency of humans to return to familiar places~\cite{song2010modelling,pappalardo2015returners}.
Inside the feedback loop, however, UR-POI's recommendations keep people returning to familiar places, making their exposure to different income groups less balanced than in their observed visits.
Offline accuracy thus rewards the same returns to familiar places that keep people segregated.

The shared order of accuracy and segregation holds in every simulation in all four metropolitan areas (Table~\ref{tab:rq1}), and extends toward more advanced algorithms (Table~\ref{tab:main}).
In Boston, the ordering by segregation also holds at every venue capacity and adoption rate tested, and when we
simulate visits to outdoor and recreation venues, such as parks, instead of food venues (Fig.~\ref{fig:wA6}a, c and Table~\ref{tab:outdoor}).

To test whether segregation stems from the travel and capacity constraints rather than from the algorithms, we also simulate random choice in Boston, which picks uniformly among the same reachable venues that still have capacity.
Random choice gives an $S_i$ of .163, below that under every algorithm (Table~\ref{tab:main}).
\emph{Experienced income segregation therefore depends on how the algorithms rank venues, not just on which venues are close enough for people to visit.}
This means that platforms, which choose the ranking, can affect how segregated their users become, with long-term implications for their opportunities to form ties across income groups.

\subsection{Distribution of the Burden across Income Groups and Neighbourhoods}
Citywide averages can conceal whether the burdens of recommendation fall disproportionately on already disadvantaged income groups and neighbourhoods.
In every metropolitan area, the lowest income quartile remains the most segregated group under every algorithm, as in the observed visits (Table~\ref{tab:app-quartiles}).
UR-POI, the most accurate algorithm, increases segregation for all four income quartiles in every metropolitan area.
In Boston, for example, the lowest income quartile's $S_i$ rises from .360 in the observed visits to .413 under UR-POI.
The most accurate algorithm thus deepens the segregation of the people who already have the fewest chances for cross-income ties.

Choosing a less segregating algorithm, however, does not necessarily reduce venue disappearance.
UserKNN and Markov, the two least segregating algorithms, learn most from other people's visits and concentrate people on the same venues.
As a result, the similarity between people's venue visits under these algorithms is 4.4 to 6.2 times that in the observed visits across the four metropolitan areas.
With visits gathered at fewer venues, more venues go unvisited.
The share of unvisited training venues is lowest under ItemKNN and highest under UserKNN in every metropolitan area (Table~\ref{tab:rq1}).
The gains in social mixing can therefore coexist with a loss of visits for local businesses.
If those losses concentrate in already less active neighbourhoods, recommendation may reinforce inequalities between places even as it reduces segregation among people.

\begin{figure}[t]
\centering
\includegraphics[width=\columnwidth]{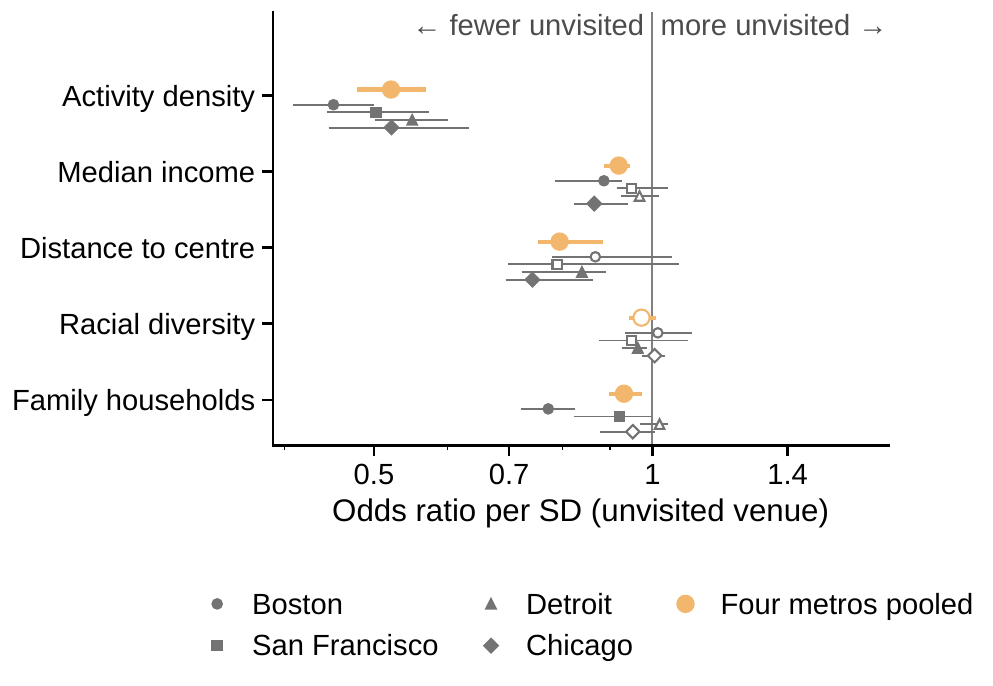}
\Description{Forest plot of odds ratios with horizontal intervals for five tract predictors and a vertical reference line at 1, with one large marker for the four metropolitan areas pooled and four small markers in different shapes, one per metropolitan area, filled where the interval excludes 1. Activity density lies far below 1 for every metropolitan area; income, distance to the centre and family households lie below 1 when pooled; for racial diversity the pooled interval crosses 1.}
\caption{Unvisited venues are more common in less active, lower-income tracts. Odds ratios that a training venue goes unvisited in the final three months, per standard deviation of each tract characteristic within its metropolitan area (activity density: training visits per km$^2$). Bars: 95\% spatial-block bootstrap intervals; filled markers: intervals that exclude 1.}
\label{fig:c5}
\end{figure}

\emph{Venue disappearance is highest in neighbourhoods that were already less active} (Fig.~\ref{fig:c5}).
We examine this pattern using a quasi-binomial logistic regression pooling the four metropolitan areas and algorithms.
The model relates each census tract's share of unvisited training venues to five tract characteristics, each standardised within its metropolitan area.
The characteristics are activity density (training visits per km$^2$), median household income, distance to the centre (the city hall of the area's main city), racial diversity (the entropy of the shares of five racial and ethnic groups) and the share of family households.
Each standard-deviation increase in a tract's log-transformed activity density is associated with a 48\% reduction in the odds that a venue goes unvisited (odds ratio .52 [.48, .57]), and this association holds in every metropolitan area.
The other characteristics have weaker associations with venue disappearance after accounting for activity density.
Venue disappearance is higher in lower-income tracts (odds ratio .92 [.89, .95]), and lower in tracts farther from the centre or with a larger share of family households.
There is no clear association with racial diversity (Fig.~\ref{fig:c5}).
Venues in less active neighbourhoods start with few visits, and a venue that receives no visits during the training window is no longer recommended after retraining.
The feedback loop thus amplifies popularity bias~\cite{mansoury2020feedback, mauro2026urban} across neighbourhoods as well as venues, concentrating the loss in less active, lower-income neighbourhoods, where small businesses rely on online platforms to attract customers~\cite{banerjee2020analyzing}.
\begin{table}[t]\centering\footnotesize\setlength{\tabcolsep}{2.5pt}
\caption{In every metropolitan area, the more accurate algorithm is the more segregating, and every algorithm leaves many venues unvisited. $S_i$: experienced income segregation; NDCG@20: offline accuracy; Unvisited: share of training venues unvisited; Real: observed visits. Averaged over five simulations, with half the 95\% interval beneath ($\pm$).}\label{tab:rq1}
\begin{tabular}{llccccc}\toprule
Metro & & UR-POI & ItemKNN & UserKNN & Markov & Real \\\midrule
Boston & NDCG@20 & .322 & .308 & .271 & .066 & -- \\
 & $S_i$ & .353 & .283 & .225 & .182 & .291 \\
\noalign{\vskip -2pt} &  & {\scriptsize\color{gray}$\pm$.001} & {\scriptsize\color{gray}$\pm$.001} & {\scriptsize\color{gray}$\pm$.005} & {\scriptsize\color{gray}$\pm$.001} & {\scriptsize\color{gray}$\pm$.005} \\
 & Unvisited (\%) & 45.8 & 21.0 & 61.8 & 47.1 & -- \\
\noalign{\vskip -2pt} &  & {\scriptsize\color{gray}$\pm$0.3} & {\scriptsize\color{gray}$\pm$0.1} & {\scriptsize\color{gray}$\pm$0.2} & {\scriptsize\color{gray}$\pm$0.2} &  \\
\addlinespace
San Francisco & NDCG@20 & .281 & .269 & .237 & .049 & -- \\
 & $S_i$ & .382 & .318 & .279 & .236 & .319 \\
\noalign{\vskip -2pt} &  & {\scriptsize\color{gray}$\pm$.001} & {\scriptsize\color{gray}$\pm$.002} & {\scriptsize\color{gray}$\pm$.003} & {\scriptsize\color{gray}$\pm$.001} & {\scriptsize\color{gray}$\pm$.005} \\
 & Unvisited (\%) & 48.7 & 29.5 & 64.8 & 45.3 & -- \\
\noalign{\vskip -2pt} &  & {\scriptsize\color{gray}$\pm$0.2} & {\scriptsize\color{gray}$\pm$0.1} & {\scriptsize\color{gray}$\pm$0.3} & {\scriptsize\color{gray}$\pm$0.1} &  \\
\addlinespace
Detroit & NDCG@20 & .316 & .299 & .268 & .060 & -- \\
 & $S_i$ & .323 & .277 & .226 & .189 & .276 \\
\noalign{\vskip -2pt} &  & {\scriptsize\color{gray}$\pm$.001} & {\scriptsize\color{gray}$\pm$.002} & {\scriptsize\color{gray}$\pm$.002} & {\scriptsize\color{gray}$\pm$.002} & {\scriptsize\color{gray}$\pm$.005} \\
 & Unvisited (\%) & 53.6 & 36.2 & 70.4 & 56.7 & -- \\
\noalign{\vskip -2pt} &  & {\scriptsize\color{gray}$\pm$0.3} & {\scriptsize\color{gray}$\pm$0.3} & {\scriptsize\color{gray}$\pm$0.2} & {\scriptsize\color{gray}$\pm$0.1} &  \\
\addlinespace
Chicago & NDCG@20 & .297 & .281 & .248 & .058 & -- \\
 & $S_i$ & .343 & .295 & .252 & .214 & .299 \\
\noalign{\vskip -2pt} &  & {\scriptsize\color{gray}$\pm$.001} & {\scriptsize\color{gray}$\pm$.001} & {\scriptsize\color{gray}$\pm$.002} & {\scriptsize\color{gray}$\pm$.000} & {\scriptsize\color{gray}$\pm$.004} \\
 & Unvisited (\%) & 54.9 & 40.6 & 70.9 & 56.8 & -- \\
\noalign{\vskip -2pt} &  & {\scriptsize\color{gray}$\pm$0.1} & {\scriptsize\color{gray}$\pm$0.2} & {\scriptsize\color{gray}$\pm$0.1} & {\scriptsize\color{gray}$\pm$0.1} &  \\
\bottomrule\end{tabular}
\end{table}
\subsection{Mitigation through a Simple Re-ranking Intervention} \label{sec:rerank}
We test whether a re-ranking intervention can lower both venue disappearance and segregation without replacing the algorithm a platform already uses.
Raising less popular venues in each person's recommendations could sustain the visits that keep them in recommendations after retraining.
Promoting less popular items is a common remedy for popularity bias in recommendations of online items such as movies~\cite{adomavicius2012improving,abdollahpouri2019managing} and of venues~\cite{delacruz2024caprifair,zhang2026give}, but for venues it has so far been evaluated only outside the feedback loop.
We test a simple form, which we call tail-boost.
Following the linear re-scoring of CAPRI-FAIR~\cite{delacruz2024caprifair}, which adds a term favouring less popular venues to the algorithm's score, tail-boost re-scores each candidate venue $\alpha$ as
\begin{equation}
  s'_\alpha = (1-\lambda)\,\hat{s}_\alpha + \lambda\,\hat{u}_\alpha, \qquad u_\alpha = \frac{1}{\log(2+n_\alpha)},
\end{equation}
where $s_\alpha$ is the algorithm's score, $n_\alpha$ the number of residents who visited $\alpha$ in the previous four months, hats denote min--max scaling, and $\lambda$ is the strength of the re-ranking.
Raising less popular venues pushes down the venues people actually visited and therefore reduces offline accuracy.
We measure this cost as the relative loss in offline NDCG@20 against the same algorithm without re-ranking.

\begin{figure}[t]
\centering
\includegraphics[width=0.85\columnwidth]{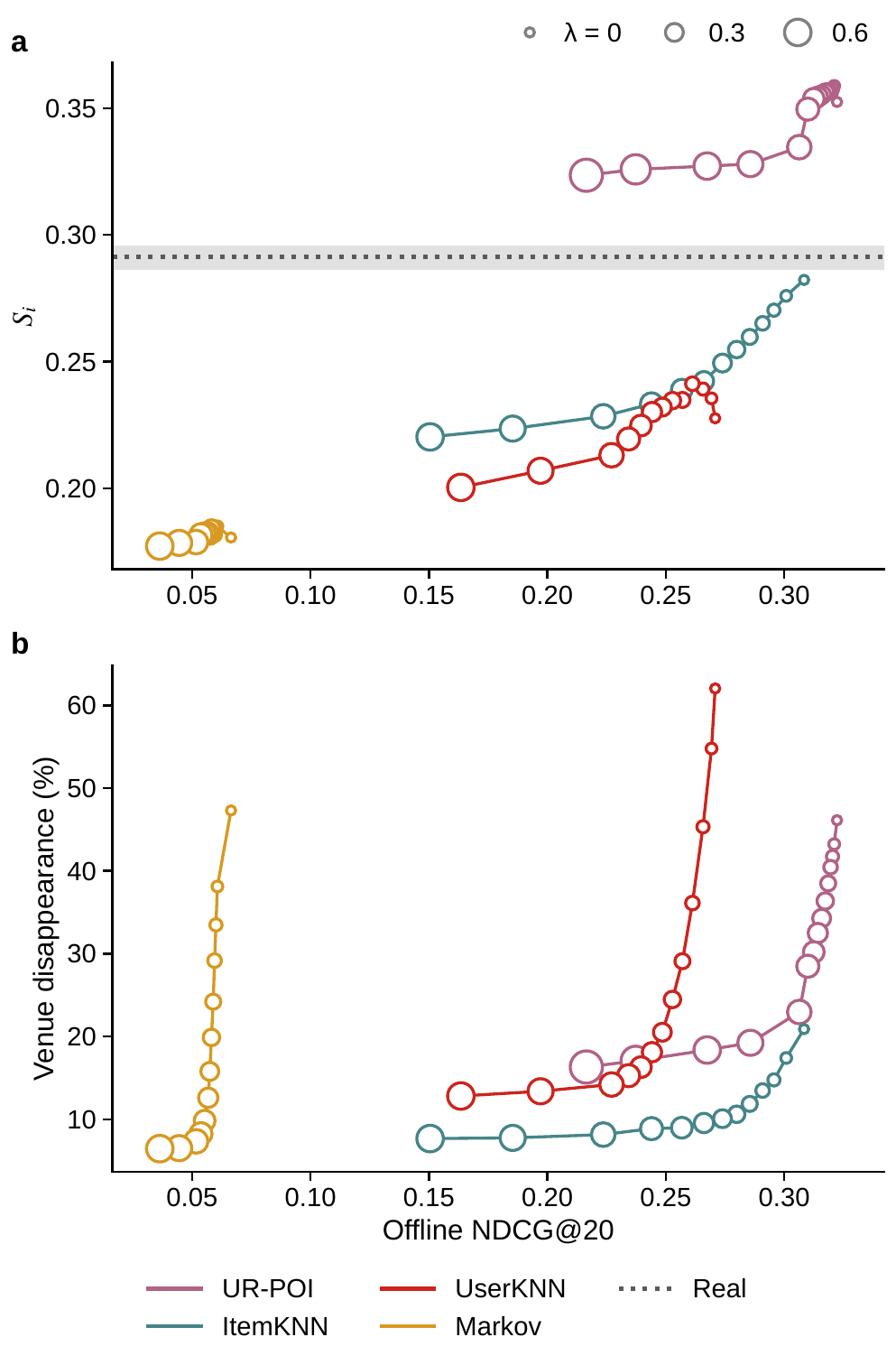}
\Description{Two stacked scatter plots sharing a horizontal axis of offline NDCG at 20. Each of four algorithms has a coloured path of hollow circles that grow with re-ranking strength from 0 to 0.6 (0.8 for UR-POI) and move left as strength grows. Top: segregation (y), with UR-POI's path above a grey observed band and the other three below it; ItemKNN's path falls from the start, while the others rise slightly before falling, Markov's only a little. Bottom: the share of venues unvisited in percent (y); every path drops steeply at first and then flattens.}
\caption{Tail-boost lowers segregation and venue disappearance at a cost in offline accuracy. Each point represents one simulation in Boston; larger markers indicate stronger re-ranking ($\lambda$). The grey band shows the observed level of experienced income segregation ($S_i$).}
\label{fig:c6}
\end{figure}

To examine how the benefits and accuracy cost vary with re-ranking strength, we simulate tail-boost in Boston at strengths from 0.05 to 0.6, and up to 0.8 under UR-POI (Fig.~\ref{fig:c6}).
Under every algorithm, most of the reduction in venue disappearance is achieved before offline accuracy falls steeply (Fig.~\ref{fig:c6}b).
By contrast, segregation falls mainly at higher strengths, except under ItemKNN, where it falls from the weakest strength tested (Fig.~\ref{fig:c6}a).
A platform can therefore obtain most of the reduction in venue disappearance at a limited loss in offline accuracy, and then decide how much more accuracy to give up for lower segregation.

We then assess how much improvement is possible within a 15\% loss in offline NDCG@20.
For each algorithm in Boston, we select the largest simulated $\lambda$ within this budget (Table~\ref{tab:rerank-budget}).
\emph{Within this budget, venue disappearance falls by 54 to 75\% under every algorithm, and segregation falls by 4 to 14\% under UR-POI, ItemKNN and UserKNN.}
Under UR-POI, segregation drops from 21\% to 13\% above the observed level.
Without replacing its algorithm, a platform can thus more than halve venue disappearance at a modest accuracy cost, with improved social mixing under three of the four algorithms.

Lower segregation could come at the cost of directing people to venues where their own income group is a small minority, places people tend to avoid when choosing where to eat~\cite{davis2019segregated,moro2021mobility}.
We therefore measure \textit{comfortableness}, the share of a person's venues where their own income quartile makes up at least half of the other visitors' time~\cite{moro2021mobility}.
Within the 15\% budget, comfortableness falls only under ItemKNN, by 7.9\% relative to no re-ranking, and rises under the other three (Table~\ref{tab:rerank-budget}).
At $\lambda = 0.6$, it changes by at most 2.4 percentage points (Table~\ref{tab:rerank}).
Tail-boost can therefore improve social mixing while largely preserving comfortableness.

Tail-boost can lower venue disappearance in two ways, as residents pick each recommended venue with probability determined by its score (Section~\ref{sec:feedback-model}).
It moves less popular venues to the top of the recommendations, and it changes the scores, which can spread people's choices more evenly across the recommendations.
To separate the two, we test a ranking-only version under ItemKNN and UserKNN in Boston at $\lambda = 0.6$.
This version keeps tail-boost's order but assigns each rank the original algorithm's score at that rank.
The chance of choosing each rank therefore stays unchanged.
This ranking-only version reduces venue disappearance at least as much as tail-boost itself under both algorithms (Table~\ref{tab:rerank}).
A platform can therefore sustain visits to less popular venues by changing which venues receive prominence, without needing residents to spread their choices more evenly across ranks.

Tail-boost raises less popular venues by the same amount for every person, but a remedy for popularity bias can also be tailored to each person's history. We test this alternative with xQuAD, a re-ranking originally designed to diversify web search results~\cite{santos2010exploiting}. Adapted to recommender systems, it reduces popularity bias in offline evaluation by matching each person's recommendations to their past share of less popular venues~\cite{abdollahpouri2019managing}.
Applied to ItemKNN and UserKNN in Boston, xQuAD has much smaller effects on venue disappearance than tail-boost, ranging from an 8.7\% reduction to a 0.9\% increase at the tested weights (Table~\ref{tab:rerank}).
Its effect is small because less popular venues make up only a small share of each person's past visits, and a list matched to this share contains few of them. Even offline, xQuAD therefore barely changes the chance of choosing a less popular venue, and under ItemKNN it lowers that chance (Appendix~\ref{app:rerank}).
This suggests that a remedy for popularity bias should be applied across people rather than matched to each person's history.

Offline evaluation, which favours the most segregating algorithm, would also reject this intervention.
It counts the accuracy cost of tail-boost but not its reductions in venue disappearance and segregation, which arise only from the visits that people make after following the recommendations.
For the same reason, it would reject a switch from UR-POI to ItemKNN, which costs only 4 to 5\% of offline NDCG@20 and lowers both segregation and venue disappearance in every metropolitan area (Table~\ref{tab:rq1}).
Platforms should therefore weigh the cost in offline accuracy against the long-term benefits of sustaining visits to local venues and creating opportunities for cross-income contact.
\begin{table}[t]\centering\footnotesize\setlength{\tabcolsep}{2pt}
\caption{Within a 15\% loss in offline NDCG@20, tail-boost lowers venue disappearance under every algorithm and $S_i$ (experienced income segregation) under three of four. In Boston, the largest simulated $\lambda$ within this budget, one simulation each. $\Delta$: \% change from no re-ranking; NDCG@20 lost: relative, averaged over people.}\label{tab:rerank-budget}
\begin{tabular}{lcccccc}\toprule
Algorithm & $\lambda$ & \shortstack{NDCG@20\\lost (\%)} & \shortstack{$\Delta$ Venue\\disappearance (\%)} & \shortstack{$\Delta$ Gini\\(\%)} & \shortstack{$\Delta S_i$\\(\%)} & \shortstack{$\Delta$ Comfort\\(\%)} \\\midrule
UR-POI & 0.55 & 11.3 & $-$58.3 & $-$17.6 & $-$7.0 & +3.2 \\
ItemKNN & 0.35 & 13.7 & $-$54.3 & $-$13.4 & $-$14.2 & $-$7.9 \\
UserKNN & 0.45 & 13.5 & $-$75.4 & $-$26.2 & $-$3.6 & +4.7 \\
Markov & 0.35 & 14.5 & $-$73.4 & $-$18.0 & +1.1 & +5.5 \\
\bottomrule\end{tabular}
\end{table}
 
\section{Discussion and Conclusion}\label{sec:discussion}
Cities create opportunity by bringing people of different incomes into the same places and by sustaining the
businesses that serve them. Location-based recommender systems that continually learn from the visits they steer can reshape both forms of opportunity, yet offline accuracy alone does not capture these collective effects.

Our simulations lead to three conclusions. First, offline evaluation is not a neutral test of how these systems affect people and places. 
It tests each algorithm only once, on visits that people made before the recommender system was in use, overlooking the effects of the human--AI feedback loop~\cite{selbst2019fairness, pedreschi2025human}. 
For places, most of the venue disappearance arises after deployment, through retraining, which offline evaluation never observes.
For people, offline accuracy rewards the same returns to familiar places that keep people segregated, and it ranks first the algorithm under which people are most segregated in the feedback loop.
Offline evaluation also rewards only predicting where people went before, and it therefore penalizes
tail-boost, the re-ranking that lowers both segregation and venue disappearance.

Second, the effects of the feedback loop fall unevenly across a city.
For places, the feedback loop amplifies popularity bias across neighbourhoods, and less active, lower-income neighbourhoods lose the most venues.
Venues without visits drop out of the data on which the recommender system is retrained, and their small businesses lose the platform they rely on to attract customers~\cite{banerjee2020analyzing}.
Across all four metropolitan areas, UR-POI (the algorithm with highest offline accuracy) further increases segregation among the lowest-income quartile, the group already most segregated in the offline data.

Third, no algorithm is best on offline accuracy, experienced income segregation, and venue disappearance at once, and the last two can be measured only inside the feedback loop.
Our results suggest that Web platforms and auditors should simulate the feedback loop with real mobility data before an algorithm is deployed, and report experienced income segregation and venue disappearance alongside offline accuracy~\cite{raji2020closing}.
With tail-boost, platforms can lower both outcomes without replacing their algorithm, at a modest cost in offline accuracy and without sacrificing residents' comfortableness.

Our results can inform targeted requests for platforms to mitigate or compensate for the effects we observe. In this respect, our methodology aligns with the EU Digital Services Act~\cite{eu2022dsa}, which requires very large online platforms to assess and mitigate systemic risks to society, including risks to fundamental rights and public well-being.

Our study has three limitations that point to future work.
First, the simulation keeps the human side of the feedback loop simple.
Residents keep the dates and durations of their observed trips, and the main setting assumes an adoption rate of 1, at which every food trip follows the recommendations.
Even so, in Boston the order of the algorithms by segregation holds at every adoption rate and venue capacity we tested (Fig.~\ref{fig:wA6}).
Second, the mobility data are anonymized to protect privacy and carry no personal incomes, and each resident's income quartile is therefore that of their home block group.
Finally, simulating the feedback loop is computationally intensive~\cite{mauro2026urban}.
We therefore tested classic algorithms with LightGCN as a neural reference, leaving heavier sequential algorithms and LLM assistants, which already expose venues unevenly across neighbourhoods~\cite{chen2026large}, for future work.

In conclusion, recommendation research has long evaluated each recommendation as an exchange between one user and one item.
In cities, every followed recommendation also changes whom other people meet and which venues stay visited, and these collective effects accumulate with each retraining.
Before deployment, we can estimate them only through simulations that realistically model where people can go, how many people each venue can serve, and whom they share it with.
With such simulations, platforms and auditors can anticipate how recommender systems on Web platforms reshape the social fabric and the local economy of cities.
{} \section{Ethics Statement}\label{sec:ethics}
Our anonymised, privacy-enhanced GPS mobility data are provided by Cuebiq and come from users who opted in to share their location under frameworks compliant with the GDPR and the CCPA.
To preserve this protection, we locate each resident's home and income only at the level of the census block group and report only aggregates over residents, venues and census tracts, never individual trajectories or named venues.
Platforms could use our simulation to select the most profitable algorithm regardless of its effects on segregation and venue disappearance.
We release it so that auditors and researchers, not only platforms, can quantify these harms.
 
\begin{acks}
E.M. and G.W. acknowledge the support from the U.S. National Science Foundation under Grants 2425023 and 2427150.
L.P. acknowledges the support from the Italian Ministry of University and Research under Directorial Decree (D.D.) no. 18010 of November 12, 2025 – FIS 3 Call - project City-AI Coevolution (CAIO).
\end{acks}

\bibliographystyle{ACM-Reference-Format}

\pagebreak
\appendix
\setcounter{figure}{0}\renewcommand\thefigure{A\arabic{figure}}\setcounter{table}{0}\renewcommand\thetable{A\arabic{table}}
\section{Appendix}\suppressfloats[t]
\subsection{Income quartiles}\label{app:quartiles}
Table~\ref{tab:app-quartiles} gives experienced income segregation $S_i$ by income quartile in each metropolitan area.
The lowest income quartile is the most segregated group under every algorithm, as in the observed visits.
\begin{table}[t]\centering\footnotesize\setlength{\tabcolsep}{3pt}
\caption{In every metropolitan area, the lowest income quartile has the highest $S_i$ (experienced income segregation) under every algorithm and in the observed visits (Real). Averaged over five simulations; Q1: lowest income.}\label{tab:app-quartiles}
\begin{tabular}{llccccc}\toprule
Metro & Quartile & UR-POI & ItemKNN & UserKNN & Markov & Real \\\midrule
Boston & Q1 & .413 & .351 & .277 & .225 & .360 \\
 & Q2 & .346 & .279 & .225 & .183 & .292 \\
 & Q3 & .330 & .254 & .203 & .164 & .257 \\
 & Q4 & .322 & .249 & .196 & .154 & .256 \\
\addlinespace
San Francisco & Q1 & .424 & .360 & .309 & .264 & .365 \\
 & Q2 & .367 & .303 & .265 & .230 & .307 \\
 & Q3 & .352 & .290 & .255 & .212 & .290 \\
 & Q4 & .384 & .321 & .289 & .239 & .315 \\
\addlinespace
Detroit & Q1 & .383 & .336 & .267 & .225 & .339 \\
 & Q2 & .327 & .285 & .231 & .196 & .284 \\
 & Q3 & .295 & .249 & .209 & .175 & .250 \\
 & Q4 & .289 & .238 & .198 & .161 & .234 \\
\addlinespace
Chicago & Q1 & .418 & .364 & .314 & .255 & .377 \\
 & Q2 & .342 & .293 & .251 & .219 & .295 \\
 & Q3 & .309 & .261 & .225 & .195 & .266 \\
 & Q4 & .303 & .261 & .220 & .186 & .258 \\
\bottomrule\end{tabular}
\end{table}
 
\subsection{The feedback loop in detail}\label{app:loop}
Table~\ref{tab:loop} gives, for each metropolitan area and algorithm, the share of unvisited venues in and out of the
loop and the changes in experienced income segregation $S_i$ and in the Gini of visits, together with the feedback loop's
share of venue disappearance and of segregation (Section~\ref{sec:conditions}).

\subsection{Algorithms and references}\label{app:all}
Table~\ref{tab:main} reports the four algorithms and the two references, LightGCN and random choice, in Boston,
together with the observed visits. LightGCN matches UserKNN in segregation but leaves far fewer venues
unvisited, and random choice gives the lowest segregation of all.

\subsection{Robustness of the order}\label{app:robust}
\begin{figure*}[b]
\centering
\includegraphics[width=\textwidth]{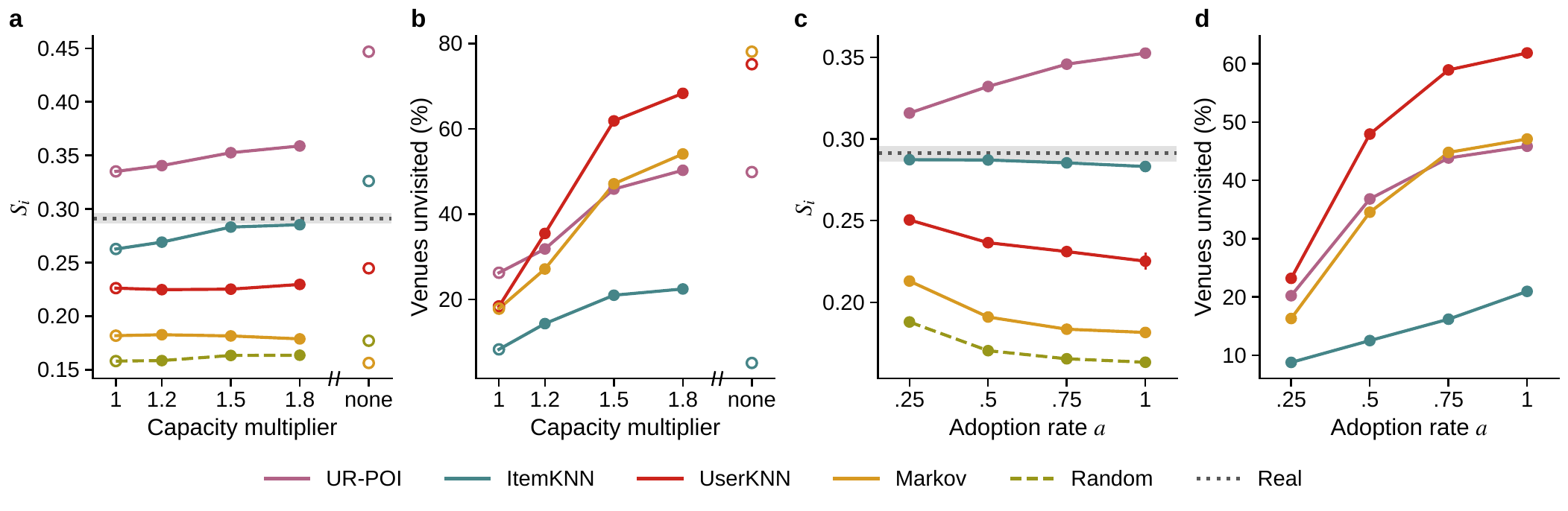}
\Description{Four line charts in one row, of segregation and the share of venues unvisited. The first two run against capacity multiplier from 1 to uncapped, the last two against adoption rate from 0.25 to 1. Each has coloured lines for four algorithms. The segregation charts add a dashed olive line for the random baseline and a grey observed band with a dotted level; the venue charts show the share of venues unvisited in percent and have no observed line. Single-simulation capacity points are hollow circles, and the uncapped point stands apart after an axis break; all other points have interval bars.}
\caption{In Boston, the order of the algorithms by $S_i$ (experienced income segregation) holds at every capacity and adoption rate tested, and ItemKNN leaves the fewest venues unvisited and UserKNN the most, except at the extreme capacities. (a, b) Capacity at adoption 1 (none: uncapped); (c, d) adoption at capacity 1.5. Last six months; hollow circles: one simulation; other points: means over five simulations with 95\% intervals; grey band: observed level. Under random choice (not drawn in b, d), almost every training venue stays visited.}
\label{fig:wA6}
\end{figure*}

In Boston, we vary venue capacity from 1.0 to uncapped and adoption from .25 to .75, and we repeat the main simulation for
outdoor and recreation venues instead of food venues (Fig.~\ref{fig:wA6} and Table~\ref{tab:outdoor}). The
order of the algorithms by $S_i$ holds in every setting, including outdoor and recreation venues. Lower adoption leaves fewer venues unvisited under every algorithm, as more
trips follow people's observed schedules.

\subsection{Re-ranking in detail}\label{app:rerank}
Table~\ref{tab:rerank} gives tail-boost at $\lambda = 0.3$ and 0.6 in Boston, with xQuAD and the ranking-only version of tail-boost
(Section~\ref{sec:rerank}). It reports offline accuracy over all recommendation decisions and over decisions
with fewer than 20 candidates, which make up most of UR-POI's decisions. The tail mass in Table~\ref{tab:rerank}
shows that xQuAD barely changes the offline choice probability on less popular venues, and under ItemKNN lowers it,
whereas tail-boost raises it sharply, in line with their effects on venue disappearance.

\begin{table*}[t]\centering\footnotesize\setlength{\tabcolsep}{2pt}
\caption{In every metropolitan area, retraining causes more than half of the venue disappearance (55--75\%) and raises the Gini of visits under every algorithm, and it pushes the algorithms apart in segregation. $\Delta$: in the loop minus out of the loop over the last six months, mean over five simulations per metropolitan area, paired by simulation, with the 95\% $t$ interval for $\Delta S_i$ ($S_i$: experienced income segregation); $\Delta$ Gini as a percentage of the out-of-loop Gini. Share: the feedback loop's share of venue disappearance or of segregation (Section~\ref{sec:conditions}). A dash marks an $S_i$ share left undefined because the observed level lies inside the in-loop 95\% interval over simulations. A negative share means that retraining moves the algorithm towards the observed level. }\label{tab:loop}
\begin{tabular}{llcccccc}\toprule
Metro & Algorithm & \shortstack{Unvisited (\%)\\in-loop} & \shortstack{Unvisited (\%)\\out-of-loop} & \shortstack{Share\\(venues)} & $\Delta S_i$ & \shortstack{Share\\($S_i$)} & \shortstack{$\Delta$ Gini\\(\%)} \\\midrule
Boston & UR-POI & 45.8 & 15.1 & 67\% & +.046 [+.044, +.047] & 75\% & +12.6 \\
 & ItemKNN & 21.0 & 5.1 & 75\% & .000 [$-$.002, +.002] & $-$2\% & +6.2 \\
 & UserKNN & 61.8 & 19.3 & 69\% & $-$.025 [$-$.031, $-$.019] & 38\% & +12.6 \\
 & Markov & 47.1 & 12.0 & 75\% & $-$.027 [$-$.029, $-$.026] & 25\% & +15.1 \\
\addlinespace
San Francisco & UR-POI & 48.7 & 19.0 & 61\% & +.046 [+.045, +.047] & 74\% & +11.9 \\
 & ItemKNN & 29.5 & 8.0 & 73\% & +.004 [+.002, +.007] & -- & +9.3 \\
 & UserKNN & 64.8 & 24.4 & 62\% & $-$.010 [$-$.013, $-$.007] & 26\% & +11.7 \\
 & Markov & 45.3 & 14.1 & 69\% & $-$.023 [$-$.025, $-$.022] & 28\% & +14.1 \\
\addlinespace
Detroit & UR-POI & 53.6 & 23.4 & 56\% & +.035 [+.034, +.037] & 76\% & +9.8 \\
 & ItemKNN & 36.2 & 11.2 & 69\% & +.003 [+.001, +.006] & -- & +7.7 \\
 & UserKNN & 70.4 & 28.5 & 60\% & $-$.021 [$-$.023, $-$.019] & 41\% & +10.9 \\
 & Markov & 56.7 & 20.0 & 65\% & $-$.029 [$-$.031, $-$.026] & 33\% & +12.7 \\
\addlinespace
Chicago & UR-POI & 54.9 & 24.5 & 55\% & +.035 [+.033, +.037] & 79\% & +10.2 \\
 & ItemKNN & 40.6 & 13.0 & 68\% & +.003 [+.001, +.004] & $-$62\% & +9.4 \\
 & UserKNN & 70.9 & 30.0 & 58\% & $-$.015 [$-$.018, $-$.013] & 33\% & +10.7 \\
 & Markov & 56.8 & 21.2 & 63\% & $-$.022 [$-$.022, $-$.021] & 26\% & +12.7 \\
\bottomrule\end{tabular}
\end{table*}
\begin{table*}[t]\centering\footnotesize\setlength{\tabcolsep}{3pt}
\caption{In Boston, experienced income segregation $S_i$ is ordered UR-POI $>$ ItemKNN $>$ UserKNN $>$ Markov in 5 of 5 simulations. Last six months, means [95\% interval]; Random and LightGCN: references; Real: observed visits. Overlap: mean similarity between people's venue visits.}\label{tab:main}
\begin{tabular}{llccccc}\toprule
Algorithm & Ranks by & $S_i$ & \shortstack{Unvisited\\(\%)} & Gini & Overlap & NDCG@20 \\\midrule
UR-POI & own visits & .353 [.352, .353] & 45.8 [45.6, 46.1] & .811 [.810, .811] & .001 [.001, .001] & .322 \\
ItemKNN & similar venues & .283 [.282, .285] & 21.0 [20.8, 21.1] & .692 [.692, .693] & .002 [.002, .002] & .308 \\
UserKNN & similar people & .225 [.220, .230] & 61.8 [61.6, 62.1] & .838 [.837, .838] & .008 [.008, .008] & .271 \\
Markov & everyone's next stop & .182 [.180, .183] & 47.1 [46.8, 47.3] & .808 [.808, .809] & .011 [.011, .011] & .066 \\
\addlinespace
Random & -- & .163 [.163, .164] & 3.0 [3.0, 3.0] & .575 [.574, .575] & .007 [.007, .007] & -- \\
LightGCN & user--venue graph & .224 [.222, .225] & 17.8 [17.5, 18.1] & .707 [.707, .708] & .007 [.007, .007] & .180 \\
\addlinespace
Real & -- & .291 [.286, .296] & -- & .629 & .002 & -- \\
\bottomrule\end{tabular}
\end{table*}
\begin{table*}[t]\centering\footnotesize\setlength{\tabcolsep}{3pt}
\caption{In Boston, the order of the algorithms by $S_i$ (experienced income segregation; UR-POI $>$ ItemKNN $>$ UserKNN $>$ Markov) also holds among outdoor and recreation venues, in 5 of 5 simulations. Main setting, last six months, means [95\% interval] over $n$ simulations; Real: observed visits.}\label{tab:outdoor}
\begin{tabular}{lccccc}\toprule
Algorithm & $S_i$ (Food) & $S_i$ (Outdoors) & Unvisited \% (Food) & Unvisited \% (Outdoors) & $n$ \\\midrule
UR-POI & .353 [.352, .353] & .416 [.415, .417] & 45.8 [45.6, 46.1] & 33.9 [33.7, 34.1] & 5 \\
ItemKNN & .283 [.282, .285] & .331 [.330, .331] & 21.0 [20.8, 21.1] & 19.9 [19.4, 20.5] & 5 \\
UserKNN & .225 [.220, .230] & .260 [.256, .265] & 61.8 [61.6, 62.1] & 59.0 [58.7, 59.3] & 5 \\
Markov & .182 [.180, .183] & .204 [.202, .205] & 47.1 [46.8, 47.3] & 53.6 [53.3, 53.9] & 5 \\
Random & .163 [.163, .164] & .181 [.181, .182] & 3.0 [3.0, 3.0] & 3.1 [3.1, 3.2] & 5 \\
Real & .291 [.286, .296] & .376 & -- & -- & -- \\
\bottomrule\end{tabular}
\end{table*}
\begin{table*}[t]\centering\footnotesize\setlength{\tabcolsep}{1.5pt}
\caption{In Boston, tail-boost leaves fewer venues unvisited under every algorithm, with changes in segregation and offline accuracy that differ by algorithm. Last six months; $\Delta$: re-ranked minus untreated, mean [95\% $t$ interval] over five simulations, paired by simulation. $\Delta$ Unvisited and $\Delta$ Gini in \% of the untreated level. Offline $\Delta$NDCG@20, weighted by decision. Comfort: comfortableness, as a level. Tail mass: offline choice probability outside the head (the venues that draw 80\% of distinct visitors). Tail-boost order: the ranking-only version of tail-boost (Section~\ref{sec:rerank}). $\lambda$ is not comparable across methods.}\label{tab:rerank}
\begin{tabular}{llccccccc}\toprule
Algorithm & Re-rank & \shortstack{$\Delta$ Unvisited\\(\%)} & $\Delta S_i$ & \shortstack{$\Delta$ Gini\\(\%)} & Comfort & $\Delta$NDCG (all) & $\Delta$NDCG ($<$20 cand.) & Tail mass ($<$20 / $\ge$20 cand.) \\\midrule
UR-POI & none & -- & -- & -- & .196 [.196, .197] & .000 & .000 & .26 / .19 \\
UR-POI & tail-boost 0.3 & $-$25.2 [$-$26.4, $-$24.0] & .002 [.001, .003] & $-$5.1 [$-$5.2, $-$5.0] & .201 [.198, .203] & $-$.009 & $-$.007 & .36 / .38 \\
UR-POI & tail-boost 0.6 & $-$59.6 [$-$60.3, $-$58.8] & $-$.025 [$-$.027, $-$.024] & $-$18.3 [$-$18.4, $-$18.1] & .199 [.198, .200] & $-$.079 & $-$.073 & .41 / .50 \\
\addlinespace
ItemKNN & none & -- & -- & -- & .167 [.166, .168] & .000 & .000 & .50 / .39 \\
ItemKNN & tail-boost 0.3 & $-$51.5 [$-$52.1, $-$51.0] & $-$.035 [$-$.037, $-$.032] & $-$12.4 [$-$12.5, $-$12.3] & .155 [.154, .156] & $-$.045 & $-$.015 & .62 / .59 \\
ItemKNN & tail-boost 0.6 & $-$63.3 [$-$64.0, $-$62.6] & $-$.064 [$-$.066, $-$.062] & $-$17.0 [$-$17.1, $-$17.0] & .144 [.142, .145] & $-$.208 & $-$.073 & .71 / .90 \\
ItemKNN & tail-boost order 0.6 & $-$68.1 [$-$68.8, $-$67.4] & $-$.051 [$-$.052, $-$.050] & $-$17.4 [$-$17.5, $-$17.3] & .151 [.149, .153] & $-$.208 & $-$.073 & .79 / .91 \\
ItemKNN & xQuAD 0.3 & $-$5.1 [$-$6.4, $-$3.8] & .000 [$-$.002, .002] & $-$1.0 [$-$1.1, $-$.9] & .167 [.164, .170] & $-$.005 & +.001 & .48 / .36 \\
ItemKNN & xQuAD 0.6 & .9 [$-$.2, 1.9] & $-$.002 [$-$.005, .000] & $-$.8 [$-$.9, $-$.6] & .166 [.163, .169] & $-$.017 & $-$.010 & .42 / .31 \\
\addlinespace
UserKNN & none & -- & -- & -- & .135 [.129, .141] & .000 & .000 & .23 / .15 \\
UserKNN & tail-boost 0.3 & $-$66.2 [$-$66.9, $-$65.6] & .008 [.003, .012] & $-$18.7 [$-$18.8, $-$18.6] & .148 [.146, .150] & $-$.023 & +.002 & .35 / .49 \\
UserKNN & tail-boost 0.6 & $-$79.5 [$-$79.8, $-$79.3] & $-$.024 [$-$.029, $-$.020] & $-$29.8 [$-$29.9, $-$29.7] & .134 [.133, .136] & $-$.143 & $-$.002 & .40 / .83 \\
UserKNN & tail-boost order 0.6 & $-$81.0 [$-$81.3, $-$80.8] & $-$.025 [$-$.030, $-$.019] & $-$30.1 [$-$30.2, $-$30.0] & .134 [.133, .136] & $-$.143 & $-$.002 & .52 / .84 \\
UserKNN & xQuAD 0.3 & $-$1.1 [$-$1.4, $-$.8] & .004 [.000, .008] & $-$.3 [$-$.4, $-$.2] & .136 [.132, .141] & $-$.001 & .000 & .23 / .17 \\
UserKNN & xQuAD 0.6 & $-$8.7 [$-$9.0, $-$8.5] & .007 [.000, .014] & $-$2.1 [$-$2.1, $-$2.1] & .142 [.139, .146] & $-$.011 & $-$.006 & .20 / .18 \\
\addlinespace
Markov & none & -- & -- & -- & .106 [.104, .108] & .000 & .000 & .85 / .21 \\
Markov & tail-boost 0.3 & $-$66.6 [$-$66.8, $-$66.4] & .001 [.000, .001] & $-$15.3 [$-$15.4, $-$15.1] & .111 [.110, .112] & $-$.014 & $-$.009 & .81 / .50 \\
Markov & tail-boost 0.6 & $-$86.5 [$-$86.7, $-$86.4] & $-$.004 [$-$.006, $-$.002] & $-$27.2 [$-$27.2, $-$27.1] & .113 [.111, .114] & $-$.049 & $-$.043 & .79 / .96 \\
\bottomrule\end{tabular}
\end{table*}
{} \FloatBarrier
\end{document}